\documentclass[12pt,a4paper]{article}
\usepackage{graphics}
\usepackage{epsf}
\newcommand{\beq}{\begin{equation}}
\newcommand{\eeq}{\end{equation}}
\newcommand{\beqar}{\begin{eqnarray}}
\newcommand{\eeqar}{\end{eqnarray}}

\newcommand{\ovar}[1]{\overline{#1}}
\begin{document}
\hfill hep-ph/0109153

\vspace*{.2cm}

\begin{center}
  {\LARGE Updates of $W_R$ effects on $CP$ angles determination
in $B$ decays }
\end{center}
\vspace*{0.1cm}
\begin{center}
{\large H.~Hayakawa}\footnote{e-mail: haya@jodo.sci.toyama-u.ac.jp}, 
{\large K.~Hosokawa}\footnote{e-mail: kaoru@jodo.sci.toyama-u.ac.jp} 
and {\large T.~Kurimoto}\footnote{e-mail: krmt@k2.sci.toyama-u.ac.jp}, 

Department of Physics, Faculty of Science,\\
Toyama University,\\
Toyama 930, Japan\\

\vspace*{1cm}
\Large{\bf Abstract} 
\end{center}
The recently observed CP violation in $B$ decay and 
$B$-$\ovar{B}$ mixing data  put constraints
on the mass of $W_R$ and the parameters of the right-handed current 
quark mixing matrix $V^R$ in $SU(2)_L \times SU(2)_R\times U(1)$ 
gauge model. It is shown that the allowed region of 
parameters are severely restricted for light $W_R$ with mass on 
the order of 1 TeV. There exist sets of parameters 
which can accommodate large CP violation as measured by Belle, 
$\sin2\phi_1|_{exp}\simeq 1$, for $M_{W_R}=1 \sim 10$ TeV.

\begin{figure}
\centerline{\resizebox{8cm}{!}{%
\includegraphics{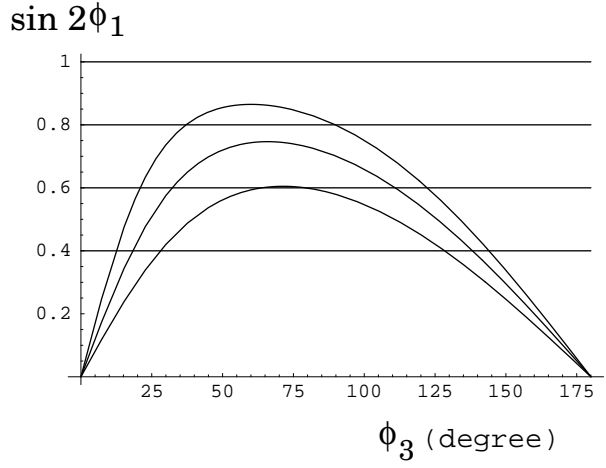}}}
\caption{
$\sin 2\phi_1$ in the 3-generation standard model.
Upper, middle and lower curves correspond to
$|V_{ub}/V_{cb}| = 0.11,~ 0.09 \mbox{ and } 0.07$, respectively.}
\label{phione}
\end{figure}

\newpage
%%%%%%%%%%%%%%%%%%%%%%%%%%%%%%%%%%%%%%%%%%%%%%%%%%%%%%%%%%%%%%%%%%%%%%%%%%%%%
The B factories at KEK and SLAC have established the existence of
CP violation in $B$ meson system by measuring the time-dependent 
CP asymmetry\cite{bcp} of neutral $B$ meson decays into 
$(c\bar c)$ meson+neutral $K^{(*)}$ meson ;
\beq
A(t) = \frac{\Gamma [\ovar{B^0}(t)\rightarrow f_{CP}] - 
             \Gamma [B^0(t)\rightarrow f_{CP}]}{
\Gamma [\ovar{B^0}(t)\rightarrow f_{CP}] + 
             \Gamma [B^0(t)\rightarrow f_{CP}]}
= -\xi_f \sin 2\phi_1 \sin (\Delta M_B t), 
\eeq
where $\xi_f$ is the CP eigenvalue of the final state.
They obtained\cite{BABAR,BELLE} 
\beq
\sin 2\phi_1 =
\left\{
\begin{array}{ll}
 0.59 \pm 0.14  \pm 0.05 & \mbox{\ \  ({\sl BABAR})}\\
 0.99 \pm 0.14 \pm 0.06 & \mbox{\ \  (Belle)}
\end{array} 
\right. .
\eeq
Let us check if the above value is consistent with the 3-generation standard model
with Kobayashi-Maskawa mechanism of CP violation\cite{KM}.
%%%%%%%
\begin{figure}[h]
  \centerline{\resizebox{5cm}{!}{\includegraphics{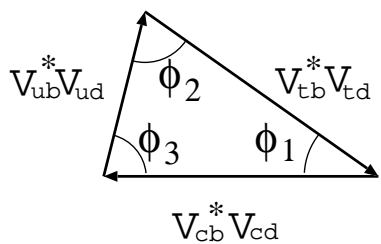}}}
\caption{Unitarity triangle}
\label{unitri}
\end{figure}
%%%%%%
With the notation of the unitarity triangle given in Fig.\ref{unitri} 
where ${V_{cb}}^*V_{cd}$ is real negative, geometrically defined 
$\sin 2\phi_1$ is given as 
\beq
\sin 2\phi_1 = 
\sin \left(2\pi -2\arg[\frac{ {-V_{tb}}^*V_{td} }{{-V_{cb}}^*V_{cd}}]
     \right)
%&=&\sin \left(-2\arg[\frac{{V_{cb}}^*V_{cd} + {V_{ub}}^*V_{ud} }{{-V_{cb}}^*V_{cd}}]\right)\\
%&=&\sin \left(-2\arg[ -1 + \left|\frac{{V_{ub}}^*V_{ud}}{{V_{cb}}^*V_{cd}} \right| 
%%e^{i\phi_3}]\right)\\
=\sin \left(2\arg[ -1 + \left|\frac{{V_{ub}}^*V_{ud}}{{V_{cb}}^*V_{cd}} 
                       \right|
    e^{-i\phi_3}]\right).
\label{p3eq}
\eeq
%%%%%%%%%
If no new physics beyond the 3-generation standard model enters 
in the measured processes of CP violation, the observed $\sin 2\phi_1$ 
should agree with the above geometrically defined one. 
Assuming that new physics does not affect the determination 
of $|V_{ud}|$, $|V_{cd}|$ and $|V_{ub}/V_{cb}|$  which are 
obtained through tree level semi-leptonic processes, the prediction of 
$\sin 2\phi_1$ in terms of $\phi_3$ by using eq.(\ref{p3eq}) and 
$|({V_{ub}}^* V_{ud})/({V_{cb}}^*V_{cd})|$ is given as Fig.\ref{phione}.
The measured result $\sin2\phi_1 > 0.4$ is consistent with 
$\phi_3 = 15^\circ \sim 145^\circ$.
The neutral $B$ meson mass difference 
$\Delta M_B$ in the standard model is also estimated as a function 
of $\phi_3$. Once $|V_{ub}/V_{cb}|$ is given, $V_{tb}{V_{td}}^*$ can be expressed 
in terms of $\phi_3$ and $|V_{ub}/V_{cb}|$ by using unitarity. Taking the ambiguity 
of hadron matrix elements to $\pm 30$\% and errors of $|V_{ub}/V_{cb}|$, 
we find the standard model is consistent with the experimental value of 
$\Delta M_B$ for $\phi_3 = 20^\circ \sim 70^\circ$.
%%%%%%%%%
%%%%%%%%%
\begin{figure}
\centerline{\resizebox{8cm}{!}{%
\includegraphics{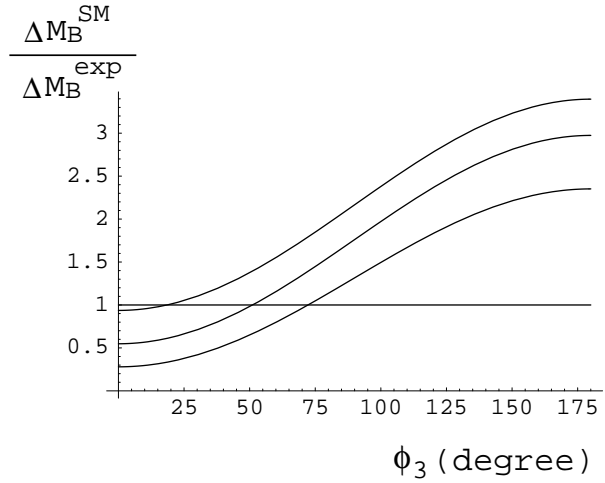}} }
\caption{$\Delta M_B^{SM}/\Delta M_B^{exp}$ in the 3-generation standard model.
Upper, middle and lower curves correspond to
$|V_{ub}/V_{cb}| = 0.07,~ 0.09 \mbox{ and } 0.11$, respectively.}
\label{dmb}
\end{figure}
%%%%%%%%%
Thus, the measured values of $\sin 2\phi_1$ are consistent with
the 3-generation standard model considering the errors at present, 
though the central value given by Belle cannot be attained in the
standard model.
If the high value by Belle is confirmed in the future experiments,
some new physics beyond the standard model is necessary.

%The observed CP asymmetry in $B$ decay gives constraints 
%on new physics beyond the standard model. 

%
%This may be a matter of statistics. 
%But if the high value is confirmed in the future experiments,
%some new physics beyond the standard model is necessary.
%
In this work we investigate $SU(2)_L\times SU(2)_R\times U(1)$ model
(L-R model)\cite{lrmd}
as a possible candidate of new physics which can give larger CP asymmetry 
in the $\sin 2\phi_1$ determination than in the standard model. We  
investigate constraints on the model, and explore the possibility 
of obtaining a high value of CP asymmetry simultaneously satisfying 
constraints by $\Delta M_B$ and $K$-$\ovar{K}$ system. 
%%%%%%%%%%%%
Some groups including 
one of the present authors have investigated L-R model and showed that 
the gauge boson coupled to right-handed charged current 
($W_R$) can affect significantly on the determination of the CP violation 
angles in $B$ decays\cite{LW,NTT,KTW,SY}. The essence is as follows;
Though $W_R$ is much heavier than the ordinary $W$ boson, some elements of 
the right-handed current quark mixing matrix $V^R$ are not necessary suppressed 
in comparison with the CKM mixing matrix elements\cite{KM,CABI}. Then 
$W_R$ can contribute significantly to some processes like $B$-$\ovar{B}$ mixing  
where ordinary $W$ boson contribution is much CKM suppressed. 

There exists a sizable contribution to $K$-$\ovar{K}$ mixing in 
L-R model from the box diagram with one $W$ and one $W_R$ 
exchange\cite{LW,OE,LS,BBP}, which allows only following 
forms of $V^R$ to avoid 
constraint from CP violation in $K$-$\ovar{K}$ mixing 
for not too heavy $W_R$ with mass of $O(1)$ TeV or less
\cite{KTW}; 
\beq
   V^R_I= \left(
       \begin{array}{ccc}
          1  & 0 & 0\\
           0 & 1 & 0 \\
           0 &  0 &  e^{i\omega}
       \end{array} \right) , \qquad
   V^R_{II}= \left(
       \begin{array}{ccc}
           0 & 1 & 0 \\
           \cos\theta_R  & 0 & -e^{i\omega}\sin\theta_R \\
           \sin\theta_R  & 0 & e^{i\omega}\cos\theta_R 
       \end{array} \right). 
  \label{eqn:rkm}
\eeq 
(If we allow fine tunings among the parameters of CKM matrix and 
those of $V^R$, there are other possibilities, which we do not consider here.)
The former, $V^R_I$, does not give significant contribution to 
either $B$-$\overline{B}$ mixing or $b$ decay, so we concentrate on 
the latter type of $V^R$ here.

The contribution to $B$-$\overline{B}$ mixing is written  as 
\beq
M_{12}^B = M_{12}^{SM} + M_{LR} + M_{RR},
\eeq 
where $M_{12}^{SM}$ is the standard model contribution, 
$M_{LR}$ from the box diagram with one $W$ and one $W_R$
exchange, and $M_{RR}$ from two $W_R$ exchange. $M_{RR}$ is 
obtained simply by exchanging $L \leftrightarrow R$ in 
the standard model contribution.  $M_{LR}$ is calculated 
from the following effective Hamiltonian\cite{NTT,mrcon,ecg};
\beq
{\cal H}^{eff}_{LR}
=\sum_{i,j=u}^t
\frac{2G_F^2M_W^2}{\pi^2}\beta_g
V^{L*}_{id}V^{R}_{ib}V^{R*}_{jd}V^{L}_{jb} J(x_i,x_j,\beta)\ 
\overline{d_R}b_L\overline{d_L}b_R + \mbox{(h.c.)},
\label{eqn:lrham}
\eeq
where $\beta_g = (g_R/g_L)^2 (M_W^2/M_{W_R}^2)$ and
$x_i= m_i^2/M_W^2$. The loop 
function is defined as  
\beq
J(x,y,\beta) \equiv \sqrt{xy}
     [(\eta^{(1)}+\eta^{(2)}\frac{xy\beta}{4})J_1(x,y,\beta) 
        -\frac{1}{4}(\eta^{(3)}+\eta^{(4)}\beta)J_2(x,y,\beta)],
\label{eqn:lrh}
\eeq
with 
\begin{eqnarray*}
  J_1(x,y,\beta) &=& \frac{x\ln x}{(1-x)(1-x\beta)(x-y)} +(x\leftrightarrow y)
     -\frac{\beta \ln \beta}{(1-\beta)(1-x\beta)(1-y\beta)} \ , \\
 J_2(x,y,\beta)&=& 
\frac{x^2 \ln x}{(1-x)(1-x\beta)(x-y)} +(x\leftrightarrow y)
     -\frac{\ln \beta}{(1-\beta)(1-x\beta)(1-y\beta)} \ ,
\end{eqnarray*}
where $\eta^{(1)-(4)}$ are QCD corrections.
We use here
$\eta^{(1)}=1.1$, $\eta^{(2)}=0.26$, 
$\eta^{(3)}=1.1$, $\eta^{(4)}=1.0$ 
as the values of QCD corrections\cite{datta}. 

\begin{figure}
    \centerline{%
\resizebox{10cm}{!}{%
\includegraphics{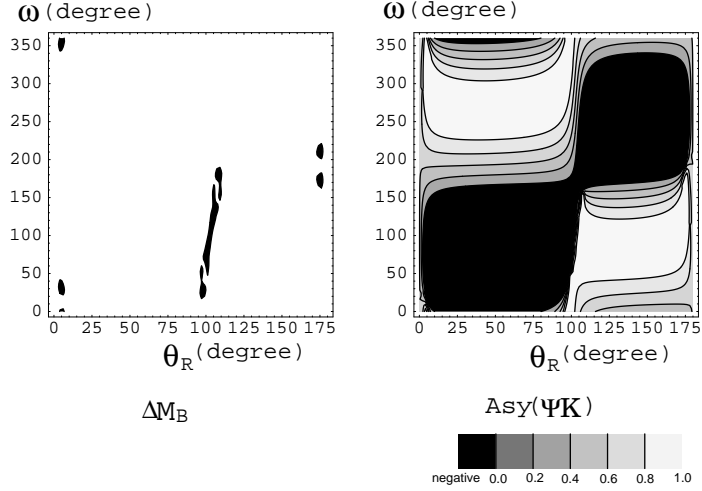}}}
    \caption{%
Allowed region by $\Delta M_B$ (left) and
$Asy(\Psi K)$ (right) for $M_{W_R} = 1$ TeV and 
$\phi_3=135^\circ$. 
Black painted regions are consistent with experimental value of $\Delta M_B$
in the left figure.}
\label{wr1t}
\end{figure}

\begin{figure}
    \centerline{ \resizebox{8cm}{!}{%
\includegraphics{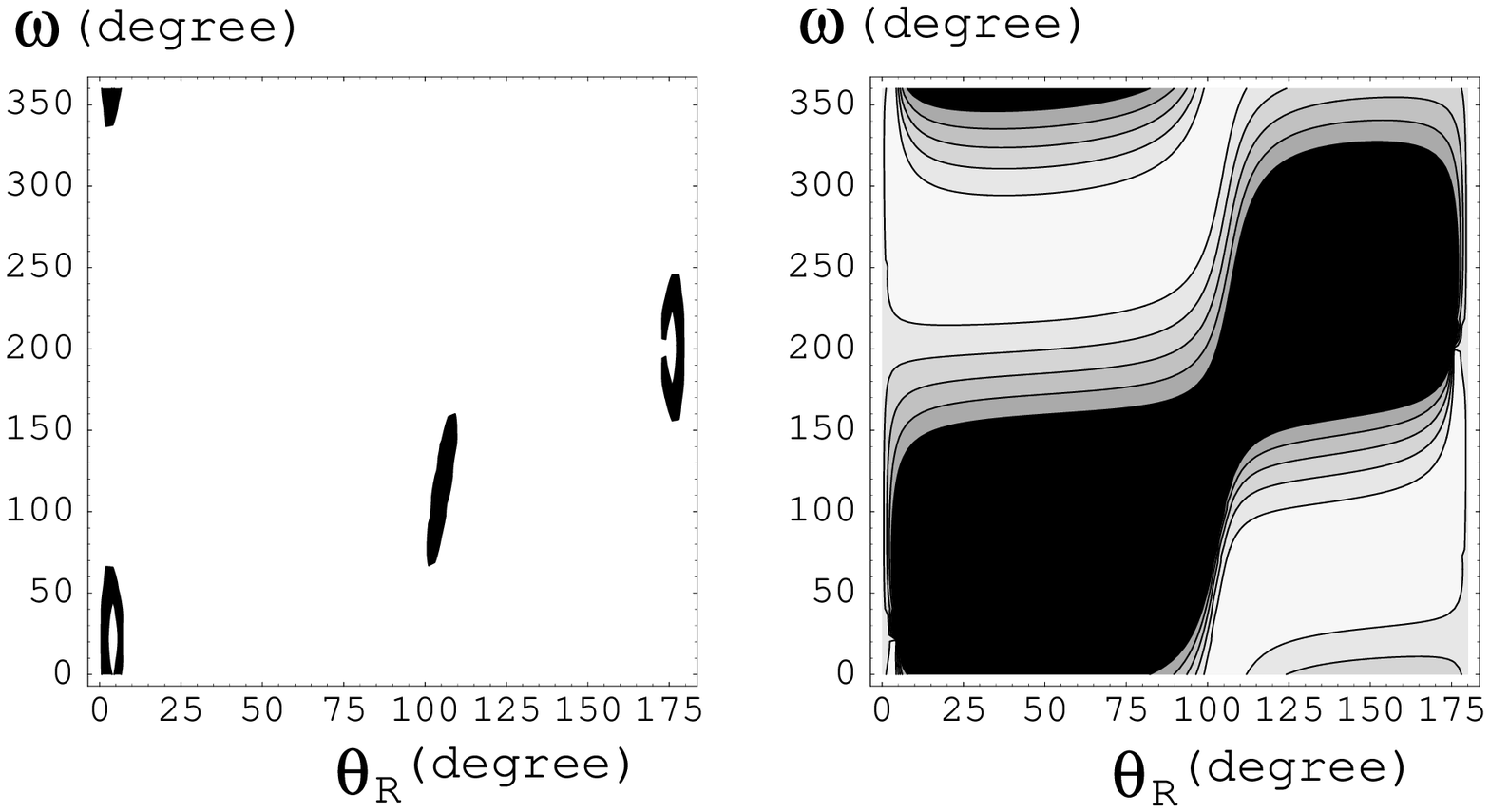}}}
\caption{%
Same as Fig.4 for
$\phi_3=90^\circ$.}
\label{wr1t90}

\centerline{ \resizebox{8cm}{!}{%
\includegraphics{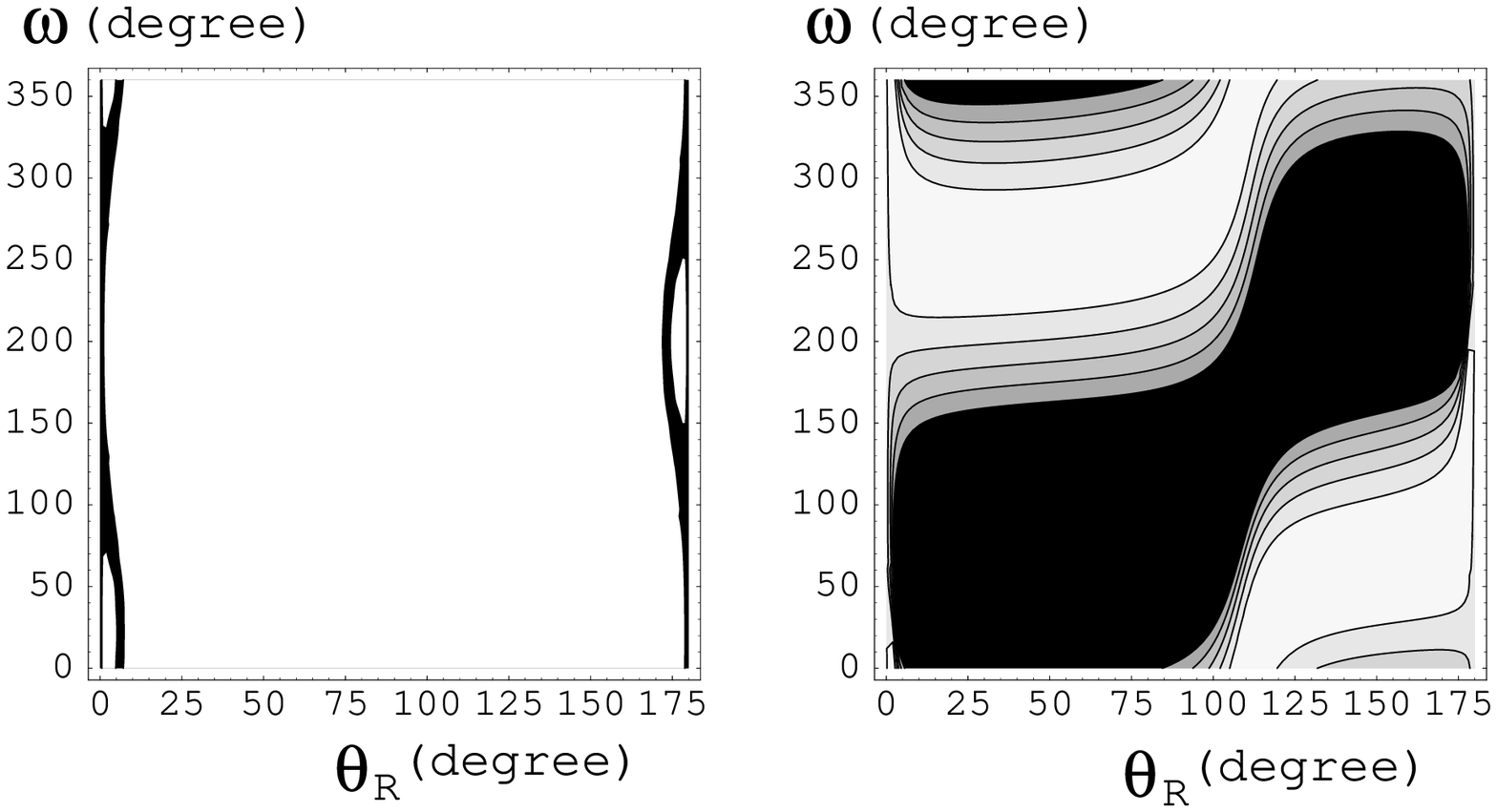}}}
\caption{%
Same as Fig.4 for
$\phi_3=45^\circ$.}
\label{wr1t45}
\end{figure}

Now we evaluate $M_{12}^B$ varying $\theta_R$ and $\omega$ in $V^R$ 
with the following inputs; 
$M(W_R) = 1 \sim 10$ TeV, $\phi_3$ in $V_{KM} = 45^\circ, 90^\circ,
135^\circ$, $|V_{ub}/V_{cb}| = 0.09$,
$f_B\sqrt{B_B} = 230$ MeV.
Then we draw allowed regions by the 
experimental values of $\Delta M_B$ allowing $\pm 30$\% 
ambiguity from errors in $f_B\sqrt{B_B}$ and $|V_{ub}/V_{cb}|$,
and 
estimate CP asymmetry in $B \rightarrow (c\bar c) + K^{(*)}$ corresponding 
to $\sin 2\phi_1$, which we call as $Asy(\Psi K)$.
First we take $M_{W_R} = 1$ TeV and $\phi_3=135^\circ$.
The allowed region and the predicted $Asy(\Psi K)$ are 
shown in Fig.\ref{wr1t}.
It can be seen that only small portions of parameter space in
$\theta_R$ and $\omega$ are allowed by $\Delta M_B$.
We fix $\theta_R = 100^\circ$ and estimate $\Delta M_B$ and
$Asy(\Psi K)$.
With $\theta_R = 100^\circ$ the CP phase $\omega$ in $V^R$ is
restricted to $30^\circ \sim 90^\circ$. If we further impose
$Asy(\Psi K) > 0.4$ from the recent measurement, $\omega$ should
be less than $60^\circ$.
It is interesting that large CP asymmetry given by Belle,
$Asy(\Psi K) \sim 1$, is possible
for $\omega = 30^\circ \sim 45^\circ$.
%%%%%%%%%%%%%%%%
%%%%%%%%%%%%%%%%
%%%%%%%%%%%%%%%%
\begin{figure}
    \centerline{ \resizebox{8cm}{!}{%
\includegraphics{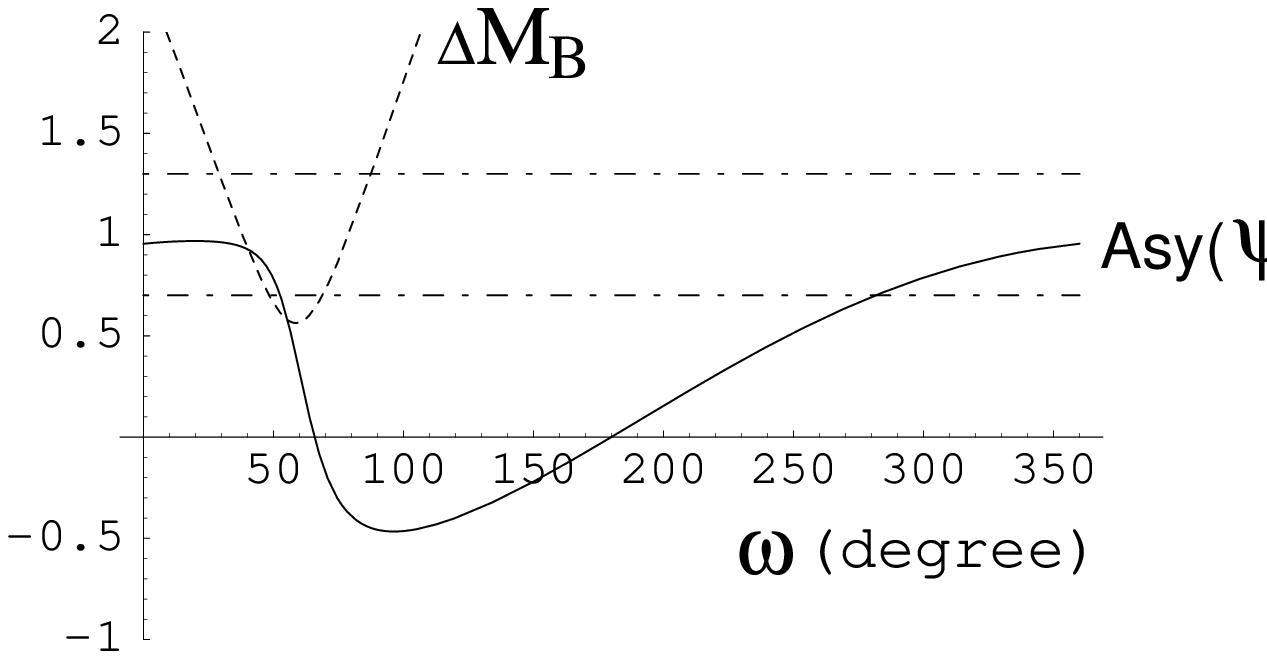}}}
\caption{%
$\Delta M_B|_{theory}/\Delta M_B|_{exp}$ and
$Asy(\Psi K)$ for $M_{W_R} = 1$ TeV,
$\phi_3=135^\circ$ and $\theta_R =100^\circ$.}
\end{figure}
%%%%%%%%%%%%%%%%
Similar figures for $\theta_R= 90^\circ$ and $45^\circ$ are
shown in Figs.\ref{wr1t90} and \ref{wr1t45}.
%%%%%%%%%%%%%%%
%%%%%%%%%%%%%%%%%
The allowed region of $\theta_R$ and $\omega$ is severely restricted.
This is because $W_R$ gives significant contribution to 
$B$-$\ovar{B}$ mixing even for $M_{W_R}\sim 1$ TeV as pointed out in 
ref.\cite{KTW}. The standard model contribution $M_{12}^{SM}$ is 
CKM suppressed by $\lambda^6$ ($\lambda \equiv |V_{us}|= 0.22$) 
while $M_{LR}$ by $\lambda^3$. Though another suppression of 
$(M_W/M_{W_R})^2$ is incorporated in $M_{LR}$, the enhancement in 
loop function and $\lambda^{-3}$ factor make $M_{LR}$ similar 
order of magnitude with $M_{12}^{SM}$.

\begin{figure}
  \centerline{\resizebox{8cm}{!}{%
\includegraphics{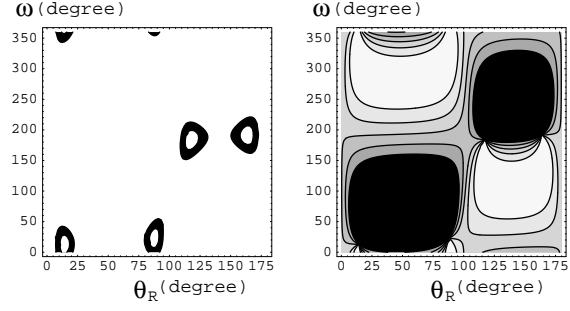}}}
\caption{%
Same as Fig.4 for $M_{W_R} = 2$ TeV and
$\phi_3=135^\circ$.}
\label{wr2t}
\end{figure}

\begin{figure}
  \centerline{\resizebox{8cm}{!}{%
\includegraphics{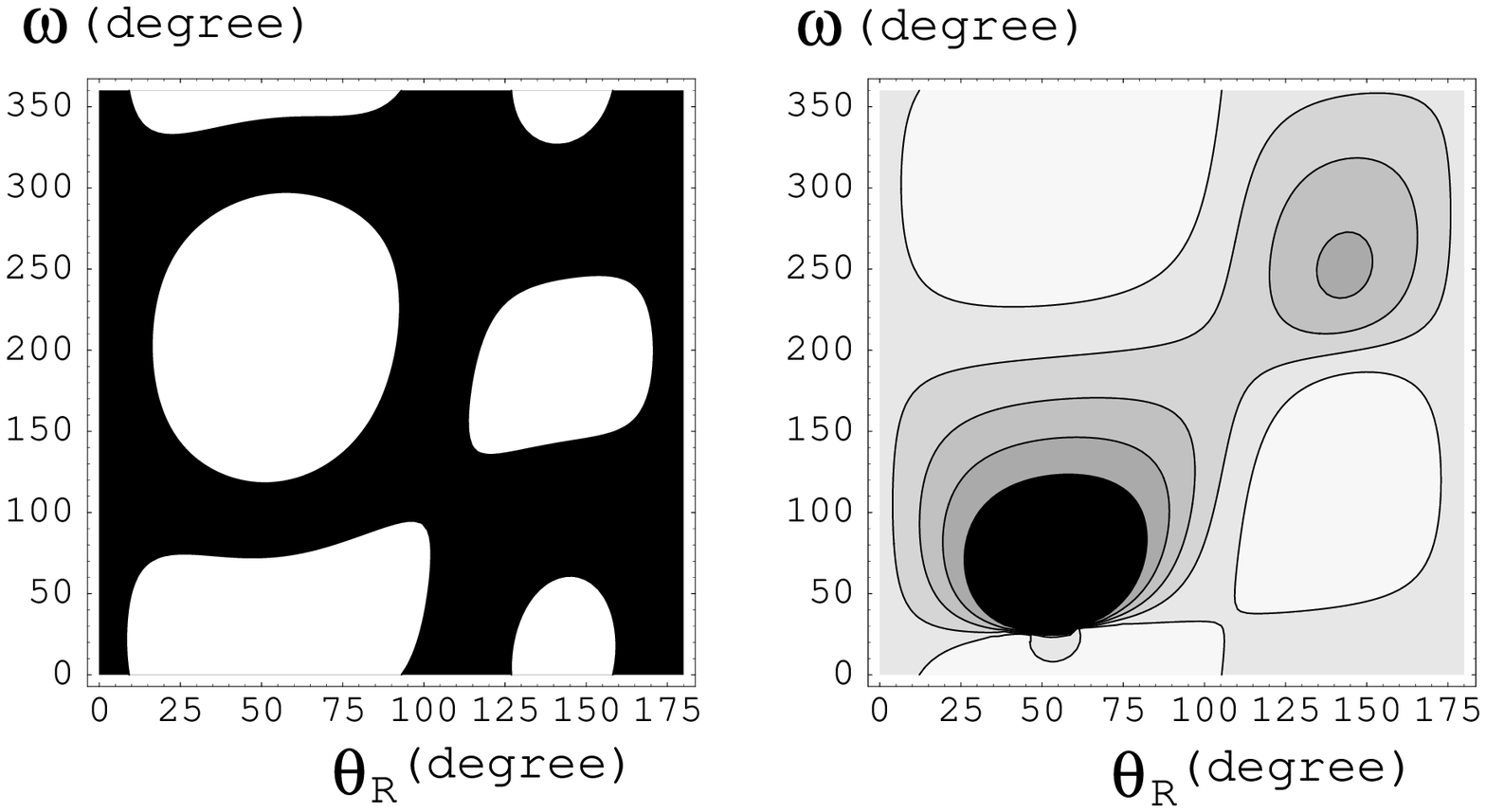}}}
\caption{%
Same as Fig.4 for $M_{W_R} = 5$ TeV and
$\phi_3=45^\circ$.}
\label{wr5t}
\end{figure}

\begin{figure}
  \centerline{\resizebox{8cm}{!}{%
\includegraphics{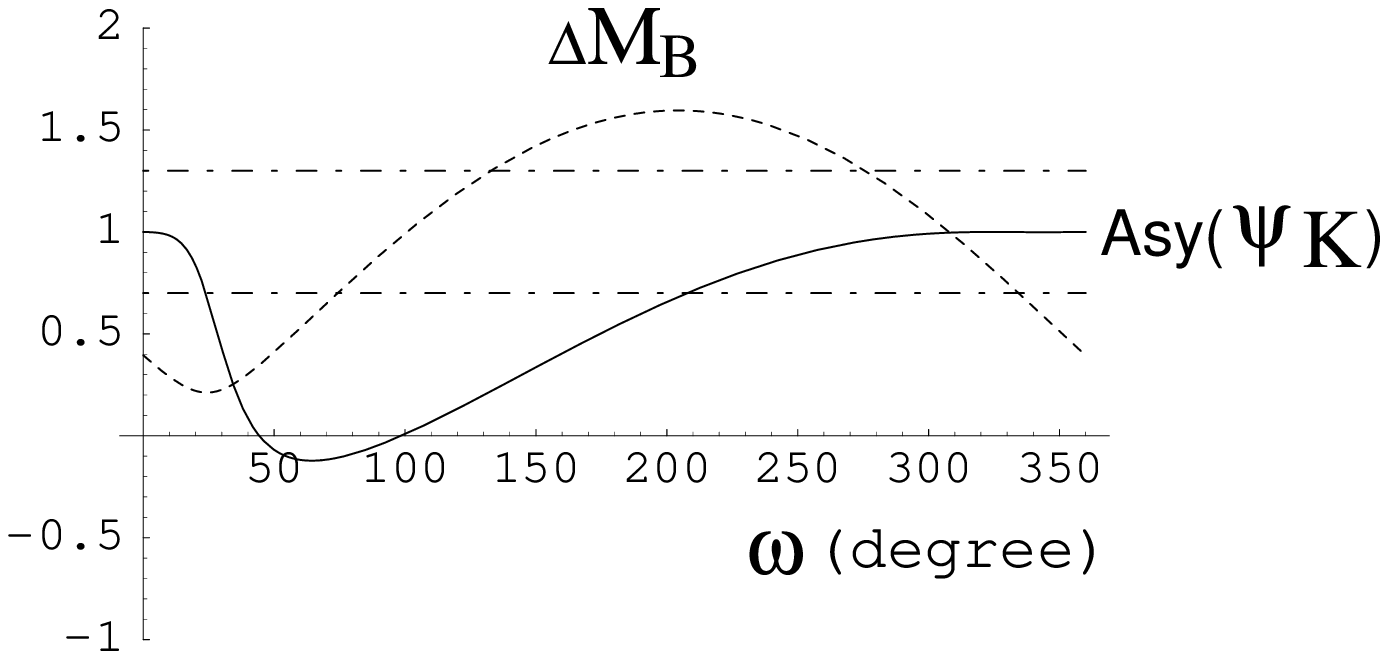}}}
\caption{%
$\Delta M_B|_{theory}/\Delta M_B|_{exp}$ and
$Asy(\Psi K)$ for $M_{W_R} = 5$ TeV,
$\phi_3=45^\circ$ and $\theta_R =30^\circ$.}
\label{wr5tp}
 \end{figure}

\begin{figure}
  \centerline{\resizebox{8cm}{!}{%
\includegraphics{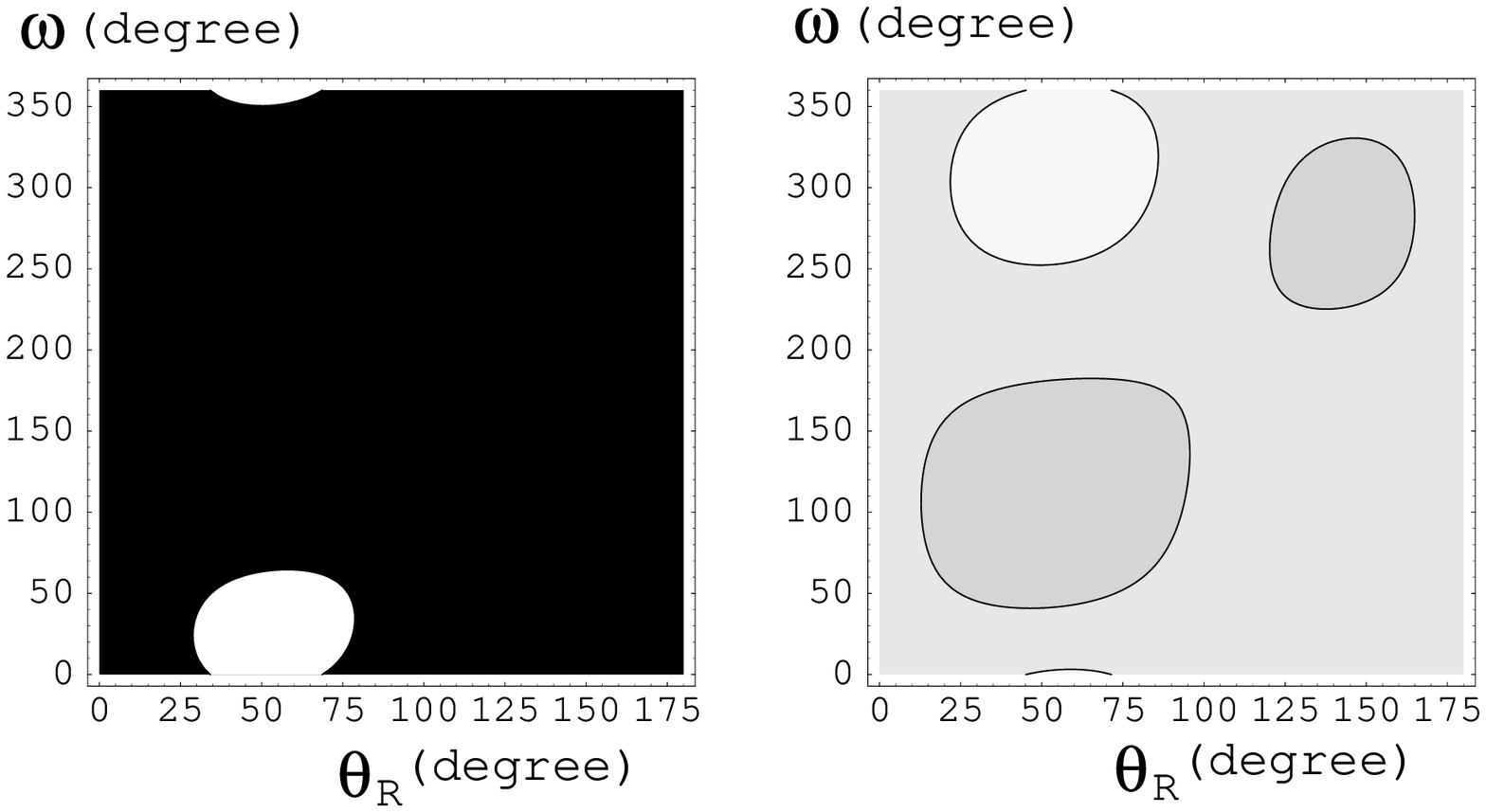}}}
 \caption{%
Same as Fig.4 for $M_{W_R} = 10$ TeV and
$\phi_3=45^\circ$.}
\label{wr10t}
\end{figure}

We have made same calculations for $M_{W_R}= 2,3,5$ and 10 TeV. The results 
are shown in Figs.\ref{wr2t}-\ref{wr5t} and \ref{wr10t}. 
The area of allowed region becomes maximal at $M_{W_R}=3$ TeV
for $\phi_3=135^\circ$.
The standard model contribution $M_{12}^{SM}$ alone cannot 
give $\Delta M_B$ consistent with the experimental data 
for $\phi_3=135^\circ$. With suitable magnitude of $W_R$ 
contribution we can give experimentally consistent value 
of $\Delta M_B$. 
Too heavy $W_R$ cannot give sufficient contribution to
compensate for $M_{12}^{SM}$. 
Similar situation occurs for
$\phi_3=90^\circ$ at more larger $M_{W_R}$.
No allowed region remains at $M_{W_R}=10$ TeV for $\phi_3=90^\circ$
and $135^\circ$.
The allowed region spreads as $M_{W_R}$ gets larger for
$\phi_3=45^\circ$ since the standard model contribution $M_{12}^{SM}$ alone
gives $\Delta M_B$ and $Asy(\Psi K)$ consistent with experimental data
for $\phi_3=45^\circ$.
It is interesting that not a small are of allowed regions which 
give large CP asymmetry remains even for heavy $W_R$. For example,
the figure for $M_{W_R} = 5$ TeV,
$\phi_3=45^\circ$ and $\theta_R =30^\circ$ is shown in Fig.\ref{wr5tp}, 
and the figure for $M_{W_R} = 10$ TeV, $\phi_3=45^\circ$
is shown in Fig.\ref{wr10t}.

%%%%%%%%%%%%%%%%%
%%%%%%%%%%%%%%%%%%
\begin{figure} 
  \centerline{\resizebox{8cm}{!}{%
\includegraphics{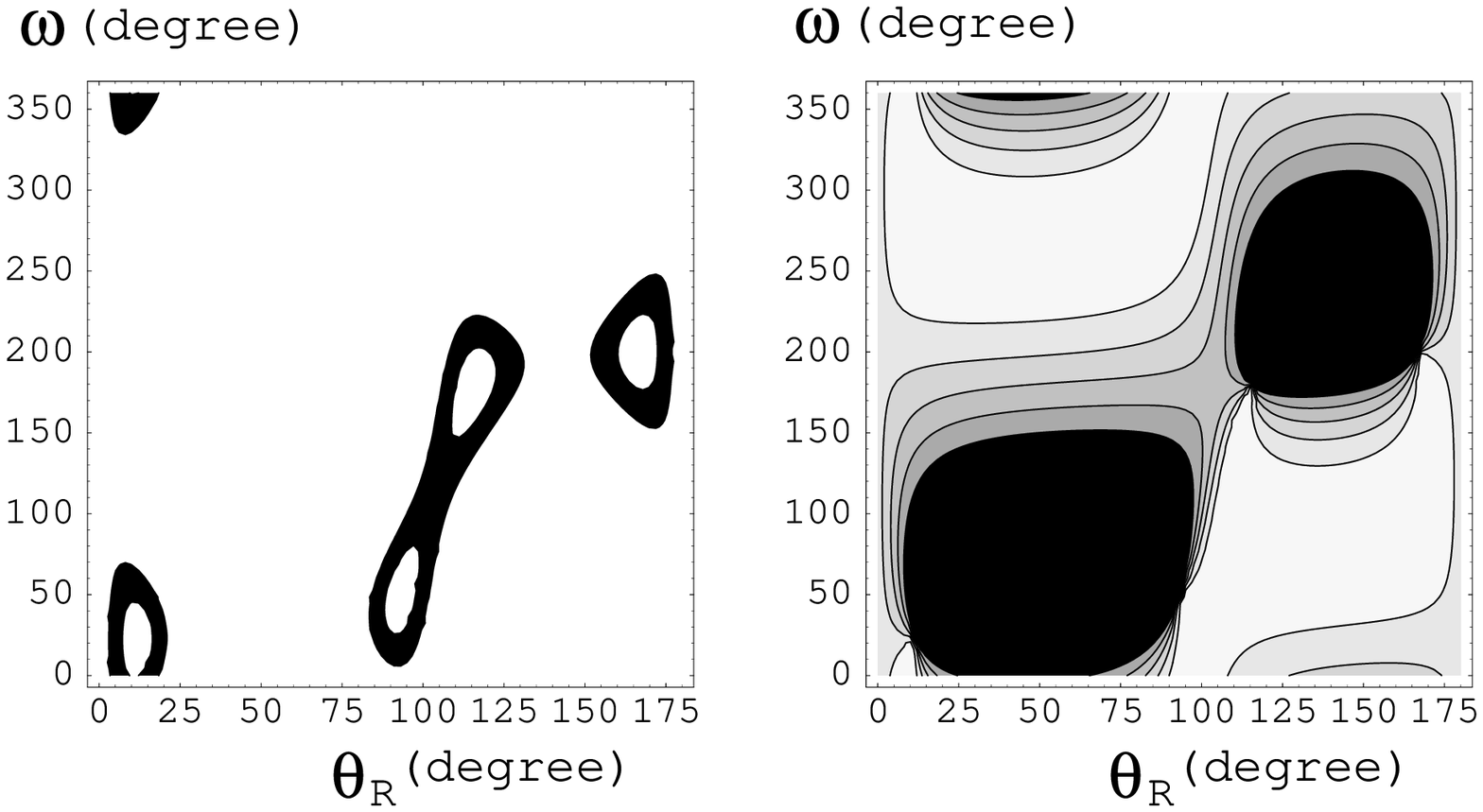}}}
\caption{%
Same as Fig.4 for $M_{W_R} = 2$ TeV and
$\phi_3=90^\circ$.}
\end{figure}
%%%%%%%%%%%%%%%%%%%%
\begin{figure}
  \centerline{\resizebox{8cm}{!}{%
\includegraphics{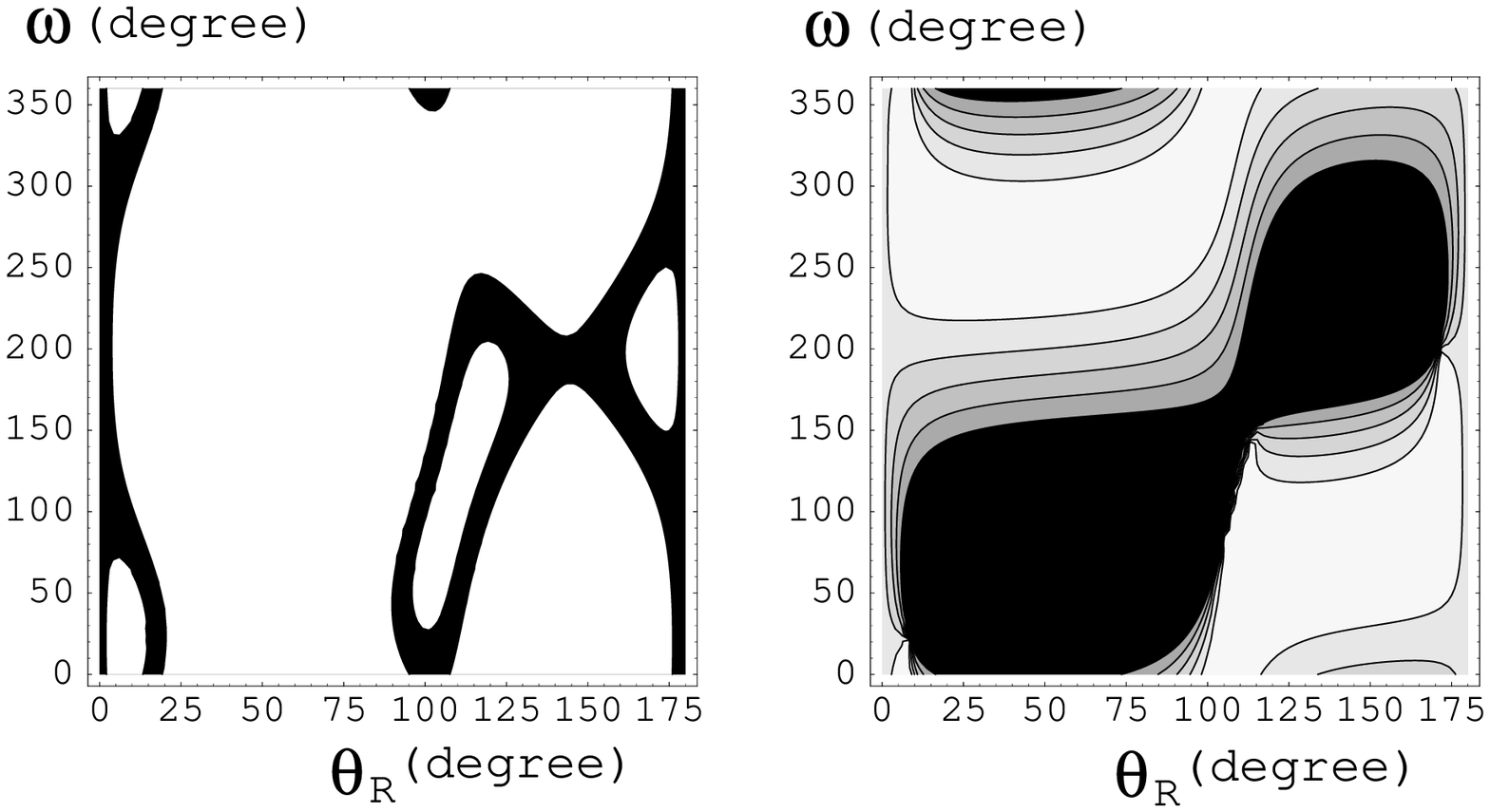}}}
\caption{%
Same as Fig.4 for $M_{W_R} = 2$ TeV and
$\phi_3=45^\circ$.}
\end{figure}
%%%%%%%%%%%%%%%%%%%%
\begin{figure}
  \centerline{\resizebox{8cm}{!}{%
\includegraphics{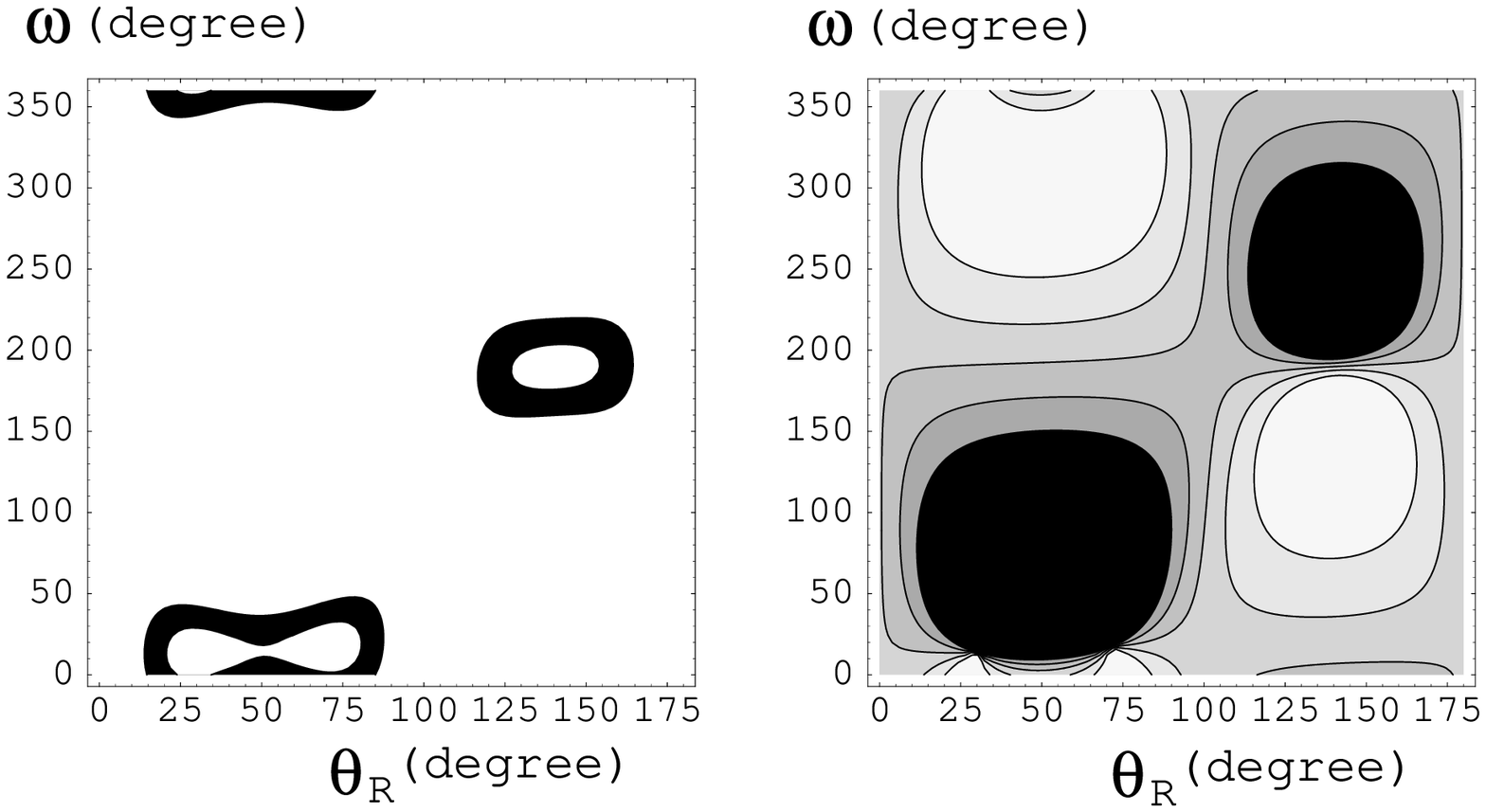}}}
\caption{%
Same as Fig.4 for $M_{W_R} = 3$ TeV and
$\phi_3=135^\circ$.}
\end{figure}
%%%%%%%%%%%%%%%%%%%%
\begin{figure}
  \centerline{\resizebox{8cm}{!}{%
\includegraphics{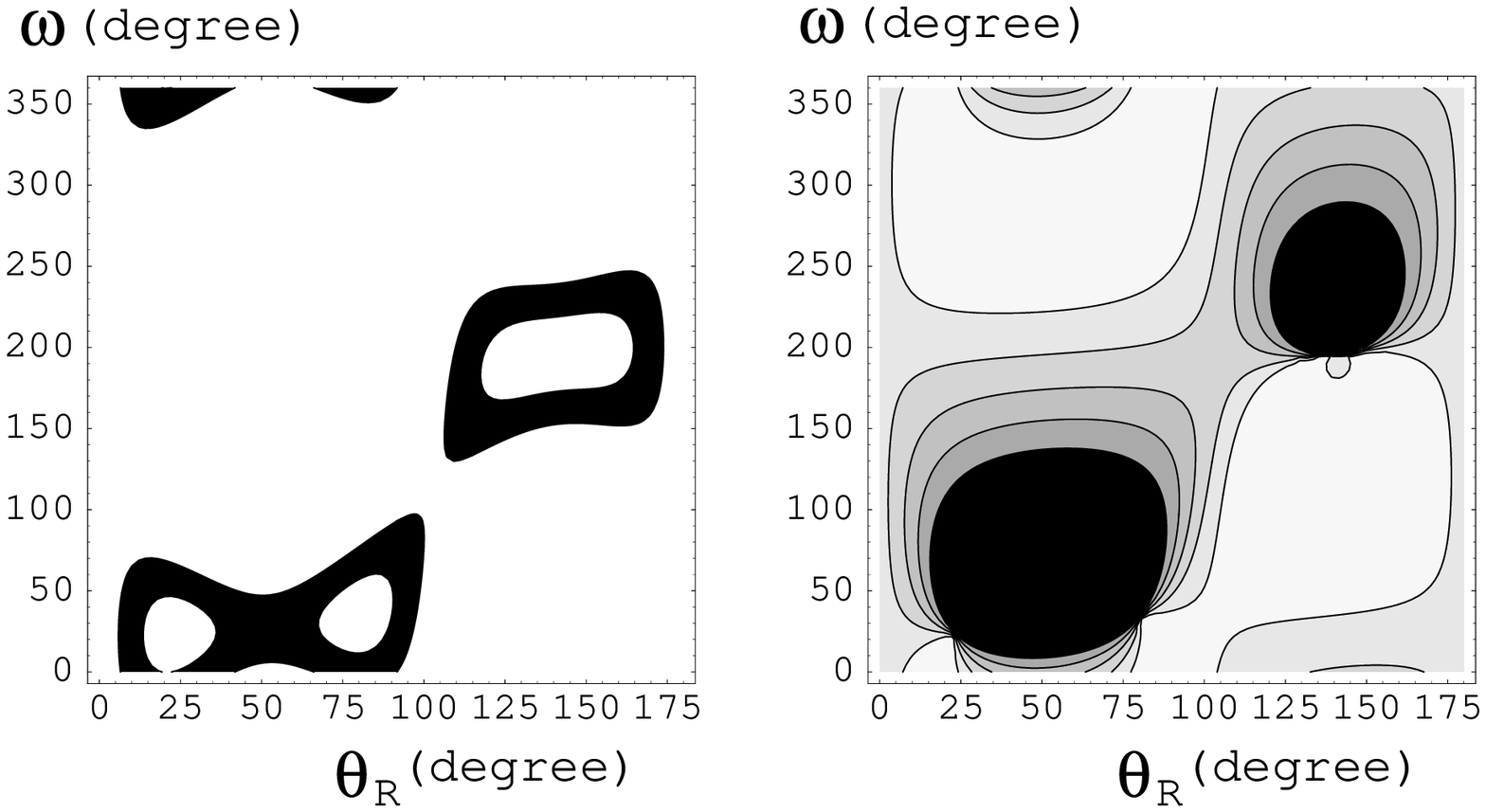}}}
\caption{%
Same as Fig.4 for $M_{W_R} = 3$ TeV and
$\phi_3=90^\circ$.}
\end{figure}
%%%%%%%%%%%%%%%%%%%%
\begin{figure}
  \centerline{\resizebox{8cm}{!}{%
\includegraphics{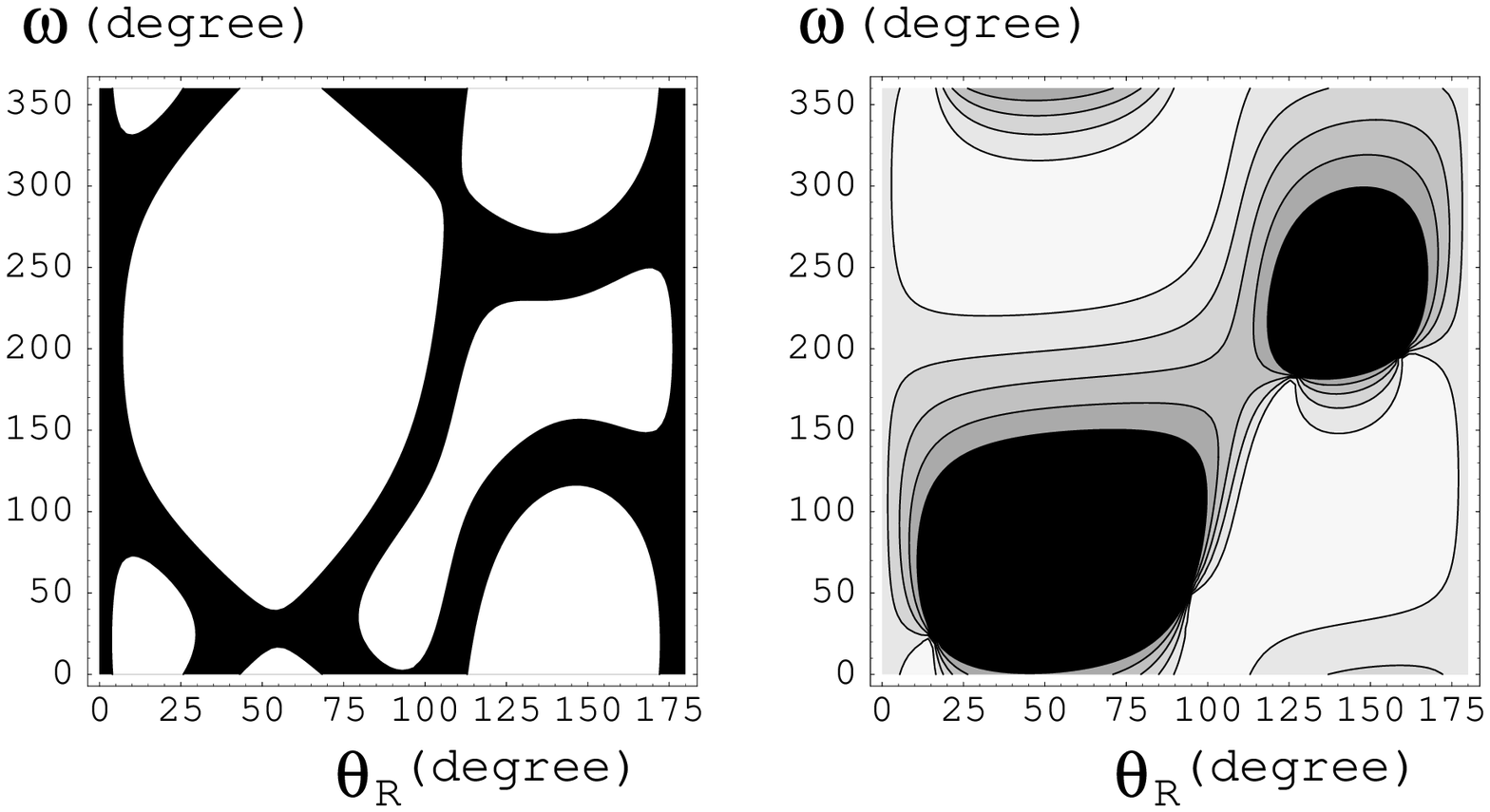}}}
\caption{%
Same as Fig.4 for $M_{W_R} = 3$ TeV and
$\phi_3=45^\circ$.}
\end{figure}
%%%%%%%%%%%%%%%%%%%%
\begin{figure}
  \centerline{\resizebox{8cm}{!}{%
\includegraphics{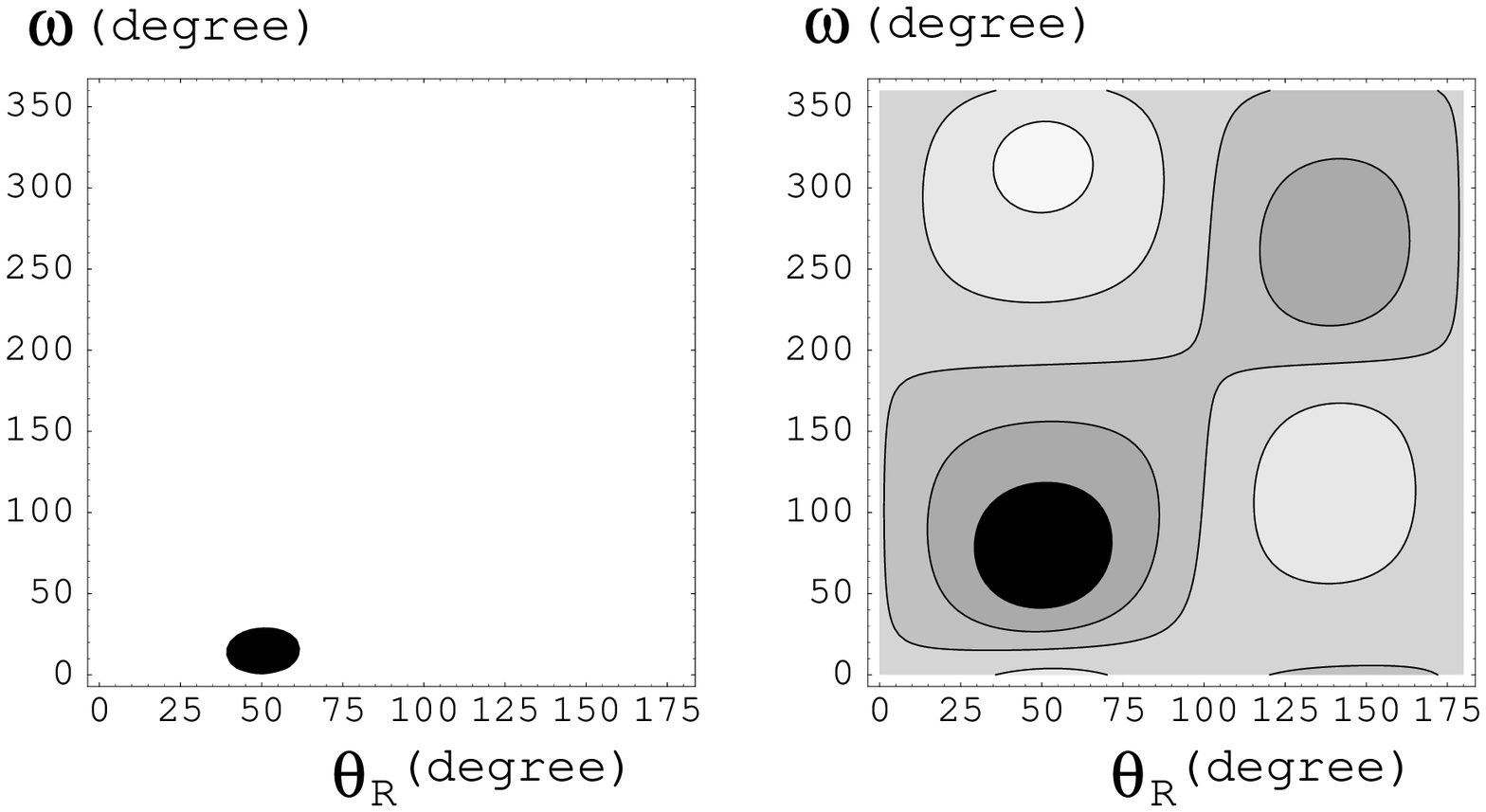}}}
\caption{%
Same as Fig.4 for $M_{W_R} = 5$ TeV and
$\phi_3=135^\circ$.}
\end{figure}
%%%%%%%%%%%%%%%%%%%%
\begin{figure}
  \centerline{\resizebox{8cm}{!}{%
\includegraphics{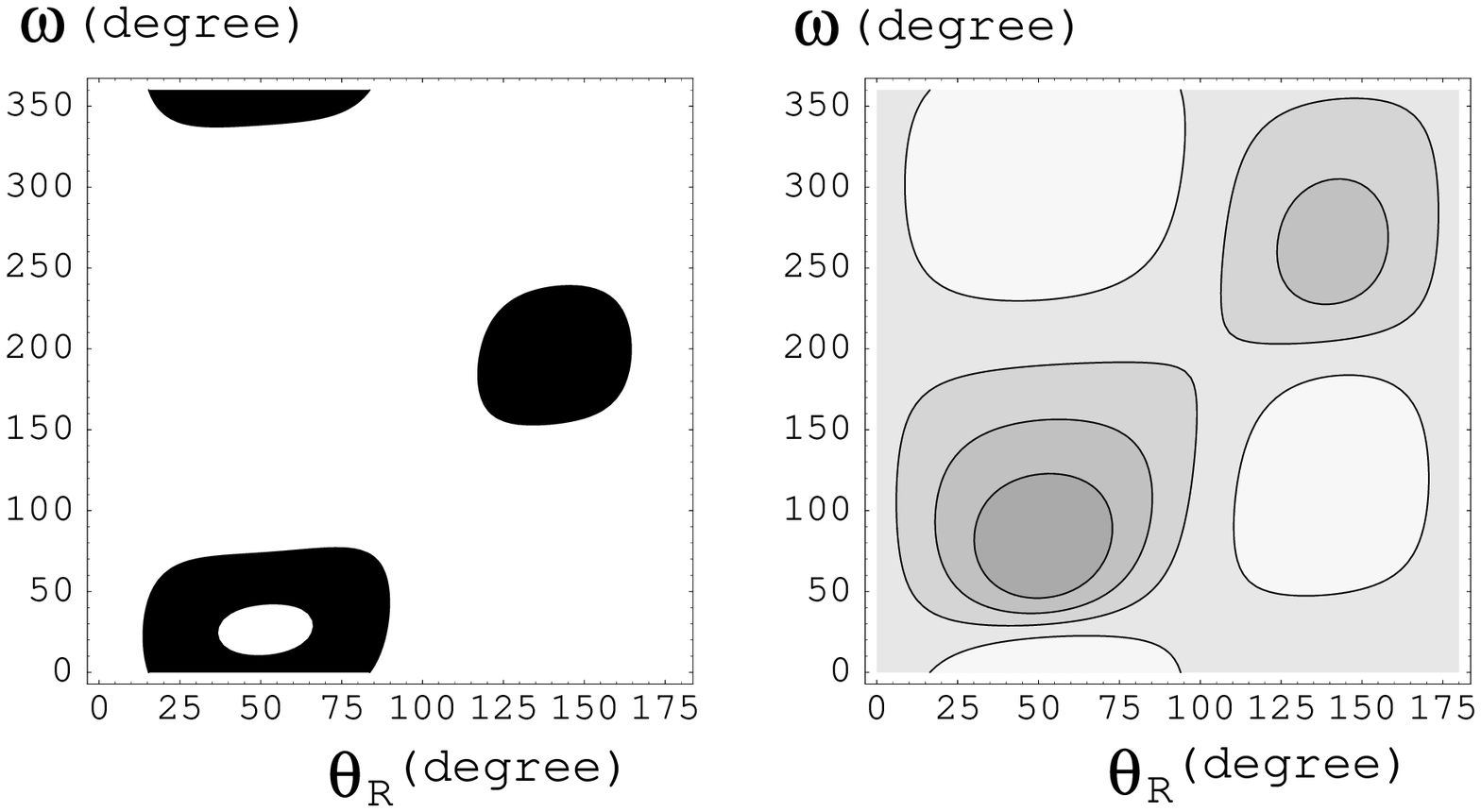}}}
\caption{%
Same as Fig.4 for $M_{W_R} = 5$ TeV and
$\phi_3=90^\circ$.}
\end{figure}
%%%%%%%%%%%%%%%%%%%%
%%%%%%%%%%%%%%%%%%%%%%%
%%%%%%%%%%%%%%%%%%%%%
%%%%%%%%%%%%%%%%%%%%%%%%%%%%%%%%%%%%%%%%%%%%%%%%5%

\begin{figure}
  \centerline{\resizebox{6cm}{!}{%
\includegraphics{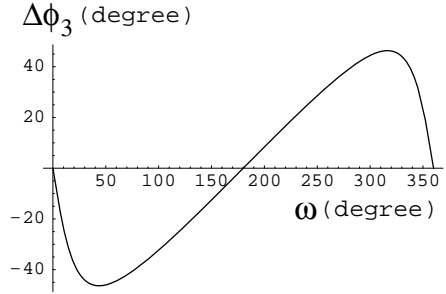}}}
\caption{%
$\Delta\phi_3$ for $M_{W_R} = 1$ TeV, 
$\phi_3=135^\circ$ and $\theta_R=100^\circ$.}
\label{wrp3}
\end{figure}

Let us comment on other CP angles, $\phi_2$ and $\phi_3$. 
$\phi_2$ is measured in the CP asymmetry in $B \rightarrow \pi\pi$ decay.
CP violation occurs through the interference among 
$B$-$\ovar{B}$ mixing, tree and penguin decays of 
$b \rightarrow u\bar ud$. $W_R$ can contribute significantly 
to $B$-$\ovar{B}$ mixing as in the case of $\phi_1$. There 
also exists contribution to $b\rightarrow d$ penguin. The ratio 
to the standard model penguin up to log loop function is given as
\beq
\frac{g_L^2}{M_W^2}{V_{tb}}^*V_{td}:
\frac{g_R^2}{M_{W_R}^2}{V^R_{tb}}^*V^R_{td} =
1: \beta_g \frac{e^{-i\omega}\sin 2\theta_R}{2{V_{tb}}^*V_{td}}. 
\eeq
The magnitude of $|\beta_g/(2{V_{tb}}^*V_{td})|$ is about 0.3 for
$M_{W_R}=1$ TeV. $W_R$ penguin is less than 10\% of 
the standard model one 
taking the allowed region of $\theta_R$ into account.
So we can neglect $b \rightarrow d$ $W_R$ penguin. 
Then the effect on $\phi_2$ is same as $\phi_1$.
For $\phi_3$ we consider the measurement by using $B^\pm \rightarrow DK$.
CP violation occurs through the interference between tree decays, 
$\bar b\rightarrow \bar c u \bar s$ and 
$\bar b\rightarrow \bar u c \bar s$ with common final 
state. $W_R$ does not contribute to $\bar b\rightarrow \bar u c \bar s$ as 
$V^R_{ub}=0$, but can affect $\bar b\rightarrow \bar c u \bar s$ decay.
\beq
   \frac{g_L^2}{M_L^2}{V_{cb}}^*V_{us}:
   \frac{g_R^2}{M_R^2}{V^R_{cb}}^*V^R_{us}  
 = 1: \beta_g \frac{ (-e^{-i\omega}\sin\theta_R)}{{V_{cb}}V_{us}}  .
\eeq
The deviation of measured $\phi_3$, $\Delta\phi_3$ from the 
standard model value for $M_{W_R}=1$ TeV, $\phi_3=135^\circ$ in CKM 
matrix and $\theta_R=100^\circ$ is given in Fig.\ref{wrp3}. 
The deviation can reach 
$-45^\circ$ for $\omega=40^\circ$
As $W_R$ gets heavier the deviation becomes small in 
proportional to $1/M_{W_R}^2$. This deviation cannot be observed 
in the measurements of $\phi_3$ in $B \rightarrow K\pi$ since 
$V^R_{us}{V^R_{ub}}^*=0$. So we can expect disagreement between 
two kinds of measurements of $\phi_3$.
%%%%%%%%%%%%%%%%%%%%
%%%%%%%%%%%%%%%%%%%%%

In conclusion, we have investigated $W_R$ effects on 
$B$-$\ovar{B}$ mixing and CP asymmetry in B decays, and 
found that $W_R$ effect is sizable even for $M_{W_R}= 1 \sim 10$ 
TeV. The experimental values of $\Delta M_B$ and CP asymmetry 
in $B \rightarrow (c\bar c)+ K^{(*)}$, $Asy(\Psi K)$,  
severely constrain the parameters 
of right-handed quark mixing matrix $V^R$. With allowed 
parameters the CP asymmetry $Asy(\Psi K)$  can 
be as large as 1 which is the central value of Belle. 
If future experiments confirms the high value of $Asy(\Psi K)$, 
fine measurements of $\phi_2$ and $\phi_3$ in various modes 
are necessary to distinguish this kind of model and other 
new physics.
\vfill

{\large\bf Acknowledgments}

This work is supported in part by Grant-in Aid for Scientific
Research from the Ministry of Education, Science and Culture of
Japan under the Grant No.11640265.

\begin{small}
%%%%%%%%%%%%%%%%%%%%%%%%%%%%%%%%

\end{small} 
\end{document}